\documentstyle[graphicx,axodraw,12pt,psfig,epsf]{article}
\makeatletter
\def\@cite#1#2{{[{#1}]\if@tempswa\typeout
{IJCGA warning: optional citation argument
ignored: `#2'} \fi}}

\newcount\@tempcntc
\def\@citex[#1]#2{\if@filesw\immediate\write\@auxout{\string\citation{#2}}\fi
  \@tempcnta\z@\@tempcntb\m@ne\def\@citea{}\@cite{\@for\@citeb:=#2\do
    {\@ifundefined
       {b@\@citeb}{\@citeo\@tempcntb\m@ne\@citea\def\@citea{,}{\bf ?}\@warning
       {Citation `\@citeb' on page \thepage \space undefined}}%
    {\setbox\z@\hbox{\global\@tempcntc0\csname b@\@citeb\endcsname\relax}%
     \ifnum\@tempcntc=\z@ \@citeo\@tempcntb\m@ne
       \@citea\def\@citea{,}\hbox{\csname b@\@citeb\endcsname}%
     \else
      \advance\@tempcntb\@ne
      \ifnum\@tempcntb=\@tempcntc
      \else\advance\@tempcntb\m@ne\@citeo
      \@tempcnta\@tempcntc\@tempcntb\@tempcntc\fi\fi}}\@citeo}{#1}}
\def\@citeo{\ifnum\@tempcnta>\@tempcntb\else\@citea\def\@citea{,}%
  \ifnum\@tempcnta=\@tempcntb\the\@tempcnta\else
   {\advance\@tempcnta\@ne\ifnum\@tempcnta=\@tempcntb \else
\def\@citea{--}\fi
    \advance\@tempcnta\m@ne\the\@tempcnta\@citea\the\@tempcntb}\fi\fi}
\makeatother

\newcommand{\gsim}{\lower.7ex\hbox{$\;\stackrel{\textstyle>}{\sim}\;$}}
\newcommand{\lsim}{\lower.7ex\hbox{$\;\stackrel{\textstyle<}{\sim}\;$}}
\newcommand{\be}{\begin{equation}}
\newcommand{\ee}{\end{equation}}
\newcommand{\bea}{\begin{eqnarray}}
\newcommand{\eea}{\end{eqnarray}}

\def\baselinestretch{1}
\begin{document}
\catcode`@=11
\newtoks\@stequation
\def\subequations{\refstepcounter{equation}%
\edef\@savedequation{\the\c@equation}%
  \@stequation=\expandafter{\theequation}
  \edef\@savedtheequation{\the\@stequation}
  \edef\oldtheequation{\theequation}%
  \setcounter{equation}{0}%
  \def\theequation{\oldtheequation\alph{equation}}}
\def\endsubequations{\setcounter{equation}{\@savedequation}%
  \@stequation=\expandafter{\@savedtheequation}%
  \edef\theequation{\the\@stequation}\global\@ignoretrue

\noindent}
\catcode`@=12
\begin{titlepage}
\title{{\bf A model of cosmology and particle physics at an
intermediate scale}}
\vskip2in
\author{
{\bf M. Bastero-Gil$^{1}$\footnote{\baselineskip=16pt E-mail: {\tt
mbg@ugr.es}}},
{\bf V. Di Clemente$^{2}$\footnote{\baselineskip=16pt E-mail: {\tt
v.di-clemente1@physics.ox.ac.uk}}}
and
{\bf S. F. King $^{3}$\footnote{\baselineskip=16pt E-mail: {\tt
sfk@hep.phys.soton.ac.uk}}}
\hspace{3cm}\\
 $^{1}$~{\small Departamento de F\'{\i}sica Te\'orica y del Cosmos,} \\
{\small Universidad de Granada, E-18071, Granada, Spain}
\hspace{0.3cm}\\
 $^{2}$~{\small Department of Theoretical Physics, University of Oxford,}
\\
{\small 1, Keble Road, Oxford OX1 3NP, U.K.}
\hspace{0.3cm}\\
 $^{3}$~{\small Department of Physics and Astronomy, University of Southampton,}
\\
{\small Highfield, Southampton, SO17 1BJ, U.K.}
\hspace{0.3cm}\\
}
\date{}
\maketitle
\def\baselinestretch{1.15}
\begin{abstract}
\noindent
We propose a model of cosmology and particle physics
in which all relevant scales arise in a natural way from an
intermediate string scale. We are led to assign the string scale to the
intermediate scale $M_* \sim 10^{13} {\rm GeV}$
by four independent pieces of physics: electroweak symmetry breaking;
the $\mu$ parameter; the axion scale; and the neutrino mass scale.
The model involves hybrid inflation with the waterfall field $N$
being responsible for generating the $\mu$ term, the
right-handed neutrino mass scale, and the Peccei-Quinn symmetry
breaking scale. The large scale structure of the Universe
is generated by the lightest right-handed sneutrino
playing the r\^{o}le of a coupled curvaton.
We show that the correct curvature perturbations
may be successfully generated providing the lightest right-handed
neutrino is weakly coupled in the see-saw mechanism,
consistent with sequential dominance.
\end{abstract}

\thispagestyle{empty}
\vspace{0.5cm}
\today
\leftline{}
\vskip-19.cm

\rightline{hep-ph/0408336}

\end{titlepage}


\setcounter{footnote}{0} \setcounter{page}{1}
\newpage
\baselineskip=20pt

\noindent

\section{Introduction}

WMAP \cite{Bennett:2003bz} has provided an unprecedented glimpse
into the early universe at the time of radiation decoupling,
which strengthens the case for a period of cosmological inflation
\cite{King:2003jw}.
With inflation becoming increasingly established, the need for
a synthesis between cosmology and particle physics becomes
ever more pressing. Such a synthesis should provide a successful
cosmological model of inflation, and cosmological perturbations
which can provide the seed of large scale structure. It should give
successful baryogenesis, for example via leptogenesis, and should
generate the required cold dark matter abundance. Ultimately it
should also explain the dark energy, but this is more ambitious
since it first requires a solution to the cosmological constant
problem, which from our present perspective seems very far away.

To achieve such a synthesis the theory should also
give a successful description of particle physics phenomena such as
right-handed neutrino masses $M_{RR}$
and a solution to the strong CP problem
such as provided for example by the Peccei-Quinn mechanism
involving an intermediate scale axion at a scale $f_a$.
The theory should also be supersymmetric, to
stabilise the hierarchy and provide flat directions
for inflation, in which case it should also provide an origin
of the Higgs $\mu$ mass parameter. Ideally the theory should
provide a complete explanation of electroweak symmetry breaking,
not only for example in terms of radiative breaking, but also
an explanation for the origin of the weak scale $M_W$.
In fact from the point of view of string theory there is only one
fundamental parameter namely the string scale $M_*$,
from which the Planck scale $M_P$ should be derived in terms
of the compactification scales. From this single scale
$M_*$ one must be able to derive all the relevant scales in physics above,
such as the axion scale $f_a$, the scales of right-handed neutrino
masses $M_{RR}$, the $\mu$ parameter, and the weak scale $M_W$,
which in the framework of supersymmetric theories is related to the
soft supersymmetry breaking masses $m_{soft}$. The successful
synthesis would therefore also be expected to provide an explanation
of all these scales in terms of a single mass scale $M_*$.

Recently we proposed a very promising model of inflation
closely related to the supersymmetric standard model
\cite{Bastero-Gil:2002xs}.
The explicit model was an extra-dimensional model \cite{Bastero-Gil:2002xs}
with an intermediate string scale
$M_*\sim 10^{13}$ GeV. This model in turn was based on an earlier model
without extra dimensions \cite{Bastero-Gil:1997vn},
and the purpose of embedding the model in extra dimensions is to
provide a natural explanation for the small Yukawa couplings and
various mass scales appearing in the model.
The lightest natural mass scale in the model turns out to be
an MeV, and this requires that the cosmological perturbations in the
model to be generated from a new mechanism which depend on
isocurvature perturbations in the slowly rolling Higgs field to be
transferred to curvature perturbations during reheating
\cite{Bastero-Gil:2002xr}. This mechanism
\cite{Bastero-Gil:2002xr} in which isocurvature
perturbations of a flat direction in hybrid inflation
become converted to curvature perturbations during reheating,
may be called a coupled curvaton scenario to distinguish it from
the weakly coupled late-decaying curvaton scenario~\cite{curvaton_lyth}.
We recently showed that in the coupled curvaton scenario,
preheating plays a crucial r\^{o}le in the conversion of the
isocurvature perturbations to the curvature perturbations
\cite{Bastero-Gil:2003tj}.

The purpose of the present paper is two-fold. Firstly by providing
a careful analysis of mass scales in the model we are led to the
remarkable conclusion that all mass scales
follow from a single input physical mass scale, namely the string scale
$M_*$ which is uniquely fixed by physics to be of order $10^{13}$ GeV.
This intermediate scale manifestly determines
the (heaviest) right-handed neutrino mass scale $M_{RR}$
as well as the Peccei-Quinn symmetry breaking or
axion scale $f_a$. In addition we show
how the assumption of a small cosmological constant fixes the
value of the supersymmetry breaking scale $F_S$ in terms of the
string scale $M_*$ and the Planck scale $M_P$. This requirement
reduces the number of free parameters in the model to just
one, the string scale $M_*$, together with the Planck scale
$M_P$ which may in principle be determined by the string dynamics
by a ratio of $M_*/M_c \sim 60$ where $M_c$ is the compactification
scale. From $M_*$ and the (in principle) string determined $M_P$
we show how the model then determines all the physical scales of
interest, as well as the dimensionless couplings.

The second main purpose of the paper is to show that
the large scale structure in the
Universe may be generated by the lightest right-handed sneutrino
playing the r\^{o}le of a coupled curvaton~\footnote{In
\cite{sneutrino_curvaton} the right-handed sneutrino plays the
r\^{o}le of the weakly coupled late-decaying curvaton.}.
Having a sneutrino rolling along a
flat direction during inflation and playing a
special r\^{o}le in determining the large scale structure of the
universe only strengthens the synthesis of particle physics
and cosmology in this approach. An oscillating sneutrino may also
allow efficient leptogenesis to take place
during the reheating process.

The layout of the remainder of the paper is as follows.
In section 2 we review the model, including the superpotential,
and then discuss qualitatively how all the
parameters which enter these potentials can be obtained from a single
mass scale $M_*$. In section 3 we discuss the potential and the
symmetry breaking aspects of the
model. In section 4 we fix the mass
scale of the theory. We also discuss the
subtle interplay between the high energy gauge group
symmetry breaking, supersymmetry and
electroweak symmetry breaking, and show how the
weak scale may be derived in terms of the string scale.
We also show how the axion scale,
the right-handed neutrino mass scale and the origin
of the $\mu$ parameter are all associated with the same string
scale. Section 5 discusses the physics during the inflationary epoch
including the special r\^{o}le played by the sneutrino as a coupled
curaton. We also discuss the physics
of preheating and reheating which is expected to give rise to
curvature perturbations. Section 7 contains our conclusions.
There are two Appendices detailing the calculation of Yukawa
couplings (Appendix A) and soft masses (Appendix B) in the presence of
extra dimensions.

\section{The model}

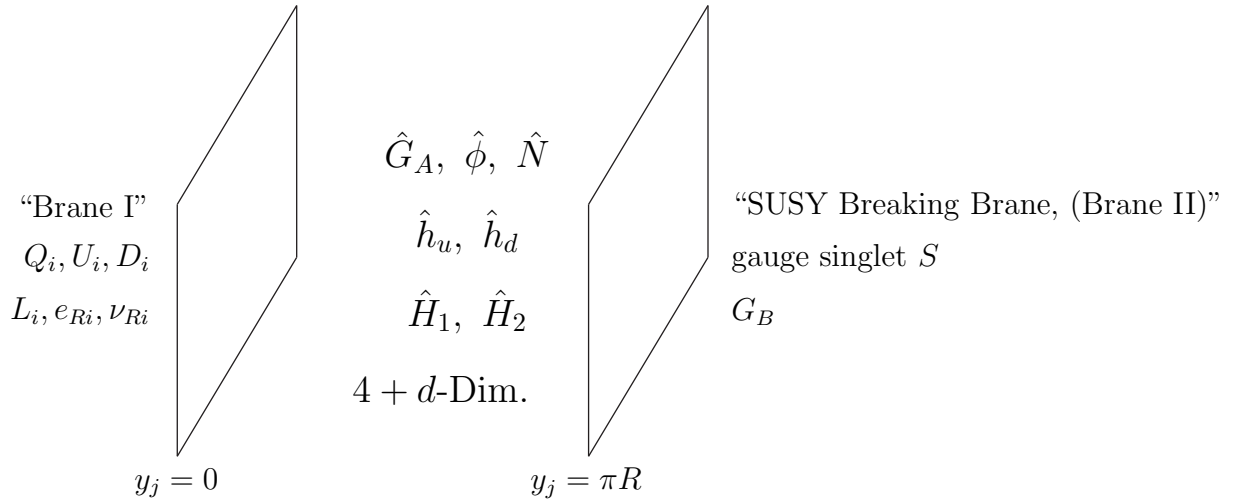
\begin{figure}[h]
\begin{center}
\begin{picture}(420,190)(0,0)
   \Line( 100, 110 )( 145, 185 )
   \Line( 100,  15 )( 145,  90 )
   \Line( 100,  15 )( 100, 110 )
   \Line( 145,  90 )( 145, 185 )
   \Text( 100,   5 )[c]{$y_j=0$}
   \Text(  90, 110 )[r]{``Brane I''}
   \Text(  90,  90 )[r]{$Q_i, U_i, D_i$}
   \Text(  90,  70 )[r]{$L_i, e_{Ri}, \nu_{Ri}$}

   \Text(210,130)[]{{\large $\hat G_A , \,\, \hat \phi, \,\, \hat N$}}
   \Text(210,100)[]{{\large $\hat h_u, \,\, \hat h_d$}}
   \Text(210,70)[]{{\large $\hat H_1, \,\, \hat H_2$}}
   \Text(200,40)[]{{\large $4+d$-Dim.}}
   \Line( 255, 110 )( 300, 185 )
   \Line( 255, 15 )( 300, 90 )
   \Line( 255, 15 )( 255, 110 )
   \Line( 300, 90 )( 300, 185 )
   \Text( 255, 5 )[c]{$y_j = \pi R$}
   \Text( 310, 110 )[l]{``SUSY Breaking Brane, (Brane II)''}
   \Text( 310, 90 )[l]{gauge singlet $S$}
   \Text( 310,  70 )[l]{$G_B$}
  \end{picture}
 \end{center}
  \caption{{\small The model showing the parallel 3-branes spatially
   separated along $d$ extra dimensions with coordinates ${\bf y}=(y_1,\dots,
   y_d)$ and a common radius $R$. The index $i$ in the matter fields
   represents the family index, $i=1,2,3$.}}
\label{prev}
\end{figure}

Let us consider two four dimensional boundaries~\footnote{From now on
  let us use the word {\it D3-brane} instead of {\it four dimensional
  boundary}.  In spite of the abuse of language in this choice,
  schematically it may represent the string conection of our model and
  also it provides a simplification of the English in this paper.}
  spatially separated along $d$ extra dimensions with a common radius
  $R = 1/M_c$. These extra dimensions are compactified on some
  orbifold that leads at least two fixed points at $\{0,\pi R\}$ where
  the two D3-branes (Brane I and Brane II) are located.
All the families of quarks and leptons are localized in one of the
fixed point ($y=0$), Brane I, while SUSY is broken by the $F$-term of
a gauge singlet field $S$ localized in the parallel brane ($y=\pi R$),
Brane II. The gauge group is $G_A \times G_B$ where $G_A$ and $G_B$
are localized in the bulk and the SUSY breaking brane
respectively. The rest of the matter are localized in the bulk, namely
the inflaton field, $\phi$, the waterfall field, $N$, the MSSM Higgs
fields $h_u$ and $h_d$, and the massive Higgs fields $H_1$, $H_2$ which
mediate the breakdown of the gauge group $G_A \times G_B$ to the
Standard Model gauge group $G_{SM} = SU(3)\times SU(2) \times
U(1)$. This is depicted in Fig. (\ref{prev}).

One of the underlying assumption of this model is that the radii of the
extra dimensions are stabilized before inflation takes place, for example,
by one of the mechanism proposed in the literature (see for
example~\cite{radii}). Done this, firstly we are going to discuss the size of
the effective four dimensional parameters (Yukawa couplings and soft
masses) in a general model with two parallel branes and then discuss
particular issues of our model.

The four dimensional superpotential of the model in Fig.
(\ref{prev}) is given by
\bea
W_4 &=& -\kappa \phi N^2 + \lambda_N
N h_u h_d + \lambda_H \phi H_1 H_2 + \lambda_{\nu ,1} \frac{N^2
(\nu_{R,1}^c)^2}{M_*} + W_{MSSM} \, , \label{superpotential}
\eea
where $W_{MSSM}$ defines the Yukawa couplings for the MSSM
\be
W_{MSSM} = y_{u}^{ij}Q_i h_u U_j + y_{d}^{ij}Q_i h_d D_j +
y_{e}^{ij}L_i h_d e_{Rj}^c + y_{\nu}^{ij} L_i h_u \nu_{Rj}^c \, .
\label{wmssm}
\ee
Following the relationship between higher
dimensional couplings ($\hat\lambda_{i,j}$) and four dimensional
one ($\lambda_{i,j}$) given in Appendix A in Eq.~(\ref{couplingd}), we get \bea
\kappa &=& \left(\frac{M_*}{M_P}\right)^3 \hat \kappa \hspace{1cm}
\lambda_N = \left(\frac{M_*}{M_P}\right)^3 \hat \lambda_N
\hspace{1cm}
\lambda_H = \left(\frac{M_*}{M_P}\right)^3 \hat \lambda_H  \nonumber \\
\lambda_{\nu} &\equiv& \lambda_{\nu ,1} \sim \left(\frac{M_*}{M_P}\right)^2 \hat \lambda_\nu \,
,\nonumber \\
y_u^{ij} & \sim & y_d^{ij} \sim y_e^{ij}
\sim
y_\nu^{ij} \sim \left(\frac{M_*}{M_P}\right) \hat \lambda_\nu \, ,
\label{our_couplings}
\eea
where for the last equation we supposed that all the higher
dimensional couplings present in $W_{MSSM}$ are equal to
$\hat\lambda_\nu$ for all families. Phenomenologically the four dimensional
MSSM couplings have to be of the order one~\footnote{The family hierarchy in the Yukawa sector can be
generated through some family symmetry very well explored in the
literature.}. Therefore it turns out that the higher dimensional coupling
is non-perturbative and it has to be $\hat\lambda_\nu \sim M_P/M_*$.

The size of the higher dimensional couplings  are completely
meaningless.  They could be either in the perturbative or in the
non-perturbative regime. From the effective field theory point of view what is
important is the size of the four dimensional couplings. However for
naturalness we impose that in the higher dimensional theory {\it all} the
couplings are of the same order
\be
\hat \kappa \sim \hat \lambda_N \sim \hat \lambda_H \sim \hat
\lambda_{\nu} \, ,
\ee
and additionaly if we also require, as experimental fact, that at least the
Yukawa coupling for the third generation (defined in $W_{MSSM}$) have to be
of order one, one gets
\bea
\kappa \sim \lambda_N \sim \lambda_H \sim \lambda \equiv
\left(\frac{M_*}{M_P}\right)^2 \, , \hspace{1.3cm}
\lambda_{\nu} \sim \left(\frac{M_*}{M_P}\right)\, .
\label{couplings_values}
\eea

Before getting into some more technical details of the model, let us
explain the
physical motivations for considering each term of the superpotential
(\ref{superpotential}).\\
{\bf 1. Inflation}: The term $\kappa \phi N^2$ will define the hybrid
inflationary potential where $\phi$ is the inflaton which slow rolls in a
semi-flat potential while the waterfall field N is set to zero. Once the
inflaton field takes some value below a critical point given by the
supersymmetric breaking sector of the model, inflation would end and the
waterfall field develops an expectation value $\langle N\rangle $. \\
{\bf 2. The $\mu$ problem:} The term $\lambda_N N h_u h_d$ will provide a
higgsino mass once inflation ends given by $\mu = \lambda_N \langle N\rangle $ like in
the Next to Minimal Supersymmetric Standard Model (NMSSM). \\
{\bf 3. Right handed neutrino masses:}
We assume that the lightest right-handed neutrino gets
Majorana mass through the non-renormalizable operator
in Eq.(\ref{superpotential}) ${\lambda_{\nu ,1}}N^2 (\nu_{R,1}^c)^2/{M_*}$,
where the Yukawa coupling is suppressed due to the fact that
the operator contains two bulk fields $N$, and is given by
$\lambda_{\nu ,1}=\lambda_{\nu } \sim {\cal O}(M_*/M_P)$. It is this lightest right-handed
neutrino $\nu_{R,1}^c$, that we shall henceforth simply refer to as
$\nu_{R}^c$, that will play the r\^{o}le of the coupled curvaton,
althought it may be completely subdominant in the see-saw mechanism,
only contributing to the
lightest physical neutrino mass $m_1$, which may be vanishingly small,
leading only to an upper bound on its Yukawa couplings.\\
{\bf 4. $G_A\times G_B$ symmetry breaking:} The vevs of $H_1$ and $H_2$
which transform under $G_A\times G_B$ as
$H_1=(\bar{R},R)$ and $H_2=(R,\bar{R})$ mediate the
breaking of the group
$G_A\times G_B$ down to the SM gauge group.
With $G_A\times G_B$ unbroken, any $F$($D$)-flat
direction would be protected against radiative corrections during inflation
arising from either Yukawa or gauge
interactions. For example, during inflation the Brane I soft masses will be
smaller than the Hubble constant (see Appendix B), this means any
$F$($D$)-flat direction would satisfy automatically the slow roll
conditions. However, well after
inflation ends the Higgses $H_1$ and $H_2$ get a vev and the Brane I soft
masses turn out to be of the order $M_{SUSY}/(4\pi)$ due to bulk particles
propagating inside a loop with $M_{SUSY}$ masses, i.e. gaugino mediation~\cite{gaugino}.

In order to specify completely the superpotential
(\ref{superpotential}) we have to impose a global $U(1)_{PQ}$
Peccei-Quinn symmetry in such a way undesirable terms like $N^3$,
$\phi^3$, $\phi h_u h_d$ and so on are forbidden. Under this
global symmetry the fields have the following charges:
\bea
&{}&
Q_N + Q_{h_u} + Q_{h_d} = 0 \, , \hspace{0.5cm} Q_\phi + 2Q_N =
0\, , \hspace{0.5cm} Q_\phi + Q_{H_1}+Q_{H_2} = 0\,,\hspace{0.5cm}
Q_N + Q_{\nu_{R,1}} = 0\, . \label{PQ} \eea
The global  symmetry,
$U(1)_{PQ}$ also forbids explicit RH Majorana neutrino masses in
the superpotential, but $B-L$ symmetry is broken by the
non-renormalizable term ${\lambda_{\nu ,1}}N^2
(\nu_{R,1}^c)^2/{M_*}$. The global symmetry is broken at the scale
of the scalar singlet VEV's releasing a very light axion and
providing as consequence an axionic solution to the CP-strong
problem. In the next section will discuss what is precisely the
axion  scale $f_a$.

\section{The potential}
Now we are ready to study in detail the scalar potential of our
model. We will write the potential along the D-flat directions  in
both Higgs sectors, $h_u=h_d=h$ and $H_1=H_2=H$, and comment later on
symmetry breaking.
Also we will take the coefficients $c_i$ for the vacuum energy
Eq. (\ref{vacuum}) and for
the soft bulk masses Eq. (\ref{bulkmass}) given in Appendix B equal to
one for simplicity,
with
\bea
V_0 &\sim& F_S^2 \, , \label{vacuum1} \\
m^2 &\equiv& \left(\frac{F_S}{M_P}\right)^2 \sim m_\phi^2 \sim m_h^2
\sim m_N^2 \sim m_H^2
\, , \label{bulkmass1} \\
A &\equiv& \frac{F_S}{M_*} = \frac{A_{\lambda_H}}{c_H}
=\frac{A_{\lambda_N}}{c_N} =
\frac{A_{\kappa}}{c_\kappa}
\label{branemass1}
\eea
With this simplifications the scalar potential for the real components
of the fields, at energy below $M_*$ looks
like
\bea
V &=& V_0 + m^2\left(\phi^2 + h^2 + N^2 + H^2\right ) +
2 \lambda A \left(c_H \phi H^2 - c_\kappa\phi N^2 + c_N N h^2 \right)
\nonumber \\
&+& \lambda^2(h^2-2\phi N)^2 + 2\lambda^2 N^2
h^2 + \lambda^2 (H^2-N^2)^2 + 2\lambda^2 \phi^2 H^2 \nonumber \\
&+& 4 \lambda_\nu^2 \frac{N^4 \tilde\nu_R^2}{M_*^2} + 4 \lambda_\nu^2\frac{N^2
\tilde\nu_R^4}{M_*^2} + 4 \lambda_\nu
\lambda\frac{N\tilde\nu_R^2}{M_*}(h^2-2\phi N) \, ,
\label{our_potential}
\eea
where the couplings $\lambda$ and $ \lambda_\nu$ are those given by
(\ref{couplings_values}), and  the first line of the above equation are the
soft susy breaking terms.

Neglecting the $m^2$-term since $m \ll A$, the global minimum of the
potential (\ref{our_potential}) is given by
\bea
\langle  \phi\rangle  &=& c_\kappa\frac{A}{4\lambda} \, ,\label{phi0} \\
\langle N\rangle  &=& c_\kappa\frac{A}{2\sqrt{2}\lambda}\,, \\
\langle H\rangle  = \langle h\rangle  &=& \langle \tilde\nu_R\rangle  = 0 \, ,
\eea
where all $c_i \approx {\cal O}(1)$. These coefficients
should  satisfy $c_H > c_\kappa/4$ ($c_N > (2-\sqrt{2})c_\kappa/8$),
in order to stabilize the vev for $h$ ($H$) at zero. On the other
hand, there is no $minimum$ with $\langle H\rangle \neq 0$ and
$\langle H \rangle \sim \langle \phi \rangle \sim \langle N \rangle$.
The solutions of the the minimization equations with $H \neq 0$ are a
maximum of the potential instead.

Using Eqs.  (\ref{couplings_values}) and
(\ref{branemass1}) the VEV are approximately
\bea
\langle \phi\rangle\sim  \langle N\rangle \sim \frac{A}{\lambda} =
\frac{F_S}{M_*}\left(\frac{M_P}{M_*}\right)^2 \, . \label{minimum1}
\eea
All parameters (couplings and mass terms) of our potential
(\ref{our_potential}) are functions of just two free
parameters\footnote{The reduced four dimensional Planck scale is fixed
at $M_P = 2.4\times 10^{18}$ GeV by gravity.}, $M_*$
and $F_S$.  However, one of them
can be eliminated by  imposing zero
cosmological constant around the global minima of the scalar
potential, $V(\langle \phi\rangle,\langle N\rangle) = 0$, and
therefore $F_S$ can be expressed as a  function of $M_*$,
\be
F_S = \frac{M_*^4}{M_P^2}\, .
\label{FS}
\ee
With this choice for $F_S$ all the
differents scales involved in our model are function of just {\it one}
scale $M_*$.
The soft masses (Eqs. ~(\ref{bulkmass1}) and
(\ref{branemass1})) can be rewritten as
\be
m^2 = M_*^2 \left(\frac{M_*}{M_P}\right)^6\, , \hspace{5mm} A = M_*
\left(\frac{M_*}{M_P}\right)^2\, .
\label{mass_A}
\ee
In general, since $M_*<M_P$ we have $m/A = M_*/M_P \ll 1$, and $\lambda
\simeq (M_*/M_P)^2 \ll 1$. Plugging
Eq. ~(\ref{FS}) into (\ref{minimum1}) we found that the VEV of the scalar
fields are degenerated at the higher dimensional cutoff scale $M_*$
\be
\langle \phi\rangle \sim   \langle N\rangle \sim M_*.
\ee

The $H_1$, $H_2$ fields will develop later a much smaller vev by a
similar mechanism to the  radiative electroweak symmetry breaking
in the Higgs sector. Extending the matter content of the model on
Brane I by two pairs of fermions $F_1$, $\bar F_1$ and
$F_2$, $\bar F_2$, in conjugate representations,
they will couple respectively to the Higgses as
$H_1F_1\bar{F_2}$ and $H_2F_2\bar{F_1}$
with Yukawa couplings of order one (same order of magnitude as the top
Yuwaka coupling). Radiative corrections due to this large Yukawa
coupling will render one of the $H_1$, $H_2$ masses negative, lifting
the D-flat direction and allowing them to get a vev. We notice that up
to this point all the matter fields on Brane I are massless. The only
massive fields are those living in the bulk.

\section{The question of scale}
In this section we shall address the numerical question:
what are the correct sizes of $m^2$ and $A$ in order
to reproduce a good phenomenology?
In other words, how large is the single mass scale $M_*$ in the theory?
The physical requirement that one of the scales $m$ or $A$ is
precisely the Electroweak scale will fix uniquely
the  value of  $M_*$,  and consequently the remaining scales.
We shall see that we are led to the conclusion that
$M_*$ must be identified with an intermediate mass scale.

\subsection{The electroweak scale}
Chiral matter do not directly feel the breaking of SUSY which takes
place in the ``hidden'' 3-brane sector. The effects for the chiral
matter of SUSY-breaking are only transmitted through the influence of
bulk fields, which are the only ones which can move into the bulk
spacetime and couple to both kinds of 3-brane sectors. As we have seen
from the Fig. (\ref{prev}) the bulk fields are the inflaton, the Higgses, the
waterfall field and gauginos belonging to the gauge group,
$G_A$. Their soft masses are equal to what we have called $m$ in
(\ref{bulkmass1}). So far we have not mentioned
gravity in this paper. It is widely believed that gravity is
propagating in the bulk in which case the gravitino mass would be
$m_{3/2} \sim m$. With this information in mind we could think that
the most natural selection for $m$ would be the Electroweak Scale,
$m_{3/2} \sim m \sim M_{SUSY}\sim$ TeV. However this is not possible
because the other scale involved, $A$,
would be $A= (M_P/M_*)$ TeV and it can be as much as $A \sim 10^3$ TeV. On
the other hand, $A$ is the scale associated with gauginos living in
the SUSY breaking brane. When the full group $G_A \times G_B$ breaks down
to the SM group, it turns out that SM gauginos would be as heavy as $10^3$
TeV which we regard as phenomenologically unacceptable.

The other possibility (the only one) is choose the scale $A$ as the
Electroweak scale. Fixing $A\sim$ TeV and using Eq. (\ref{mass_A}) we have that
the scale $M_*$ (the fundamental scale in higher dimensions) has to be
\be
M_* \approx 10^{13} {\rm GeV} \, ,
\label{M*}
\ee
As a consequence the SUSY breaking scale is $F_S^{1/2} \sim 10^8$ GeV
and the $m$-term is $m \sim 10$ MeV.  In the next section we will see
that $m$ will give us the inflaton mass. Using Eq. ~(\ref{hubble})
in Appendix B, the Hubble expansion parameter during inflation turns
our to be of the order MeV.

Some of the phenomenological benefits of an intermediate scale have been
noted in~\cite{benakli,abel}.  Below we carefully examine these issues
relevant to the present model.

\subsection{The $\mu$-scale}

A problem of the Minimal version of the Supersymmetric
Standard Model (MSSM) is why the $\mu$ term which is a supersymmetric mass
has to be of the same order of the soft terms, as required to get an
acceptable phenomenology. In the other words, what is the origin of the
$\mu$-scale?. There are many solutions to this problem~\footnote{ Note that
the Giudice-Masiero mechanism \cite{masiero} presents also a solution of
the $\mu$ problem within the MSSM by generating the $\mu$ term via a
non-minimal Kahler potential.}, for example in the
Next to Minimal model (NMSSM) the $\mu$-parameter is replaced by a
trilinear coupling involving an extra field $N$ and the higgsses, $\lambda N
H_u H_d$. Once $N$ gets a vev, the $\mu$-term is generated.

The solution of the $\mu$-problem in our model relies on the same
mechanism as in the NMSSM. However,
there are many features that make our model different to the usual NMSSM
model. The usual NMSSM involves a term like $\kappa N^3$ in the
superpotential so that the model has an exact $Z_3$
symmetry~\cite{fayet,nmssm} which
is broken at the weak scale (at the scale of
$\langle N\rangle$) leading to a serious domain wall
problem\footnote{See for example \cite{tamvakis} for an alternative
  solution to the
  domain wall problem based on a ${\cal Z}_2$
  $R$-symmetry.}~\cite{zeldovich}. In our
model a global $U(1)_{PQ}$ symmetry forbids such cubic terms
so there is no domain wall problem.
As we will see in the next subsection the global $U(1)_{PQ}$ symmetry
is linked with the solution to the
CP - strong problem. In fact the singlet field $N$ in our model plays
three r\^{o}les. It switches on the $\mu$-term once it gets a vev
\be
\mu = \lambda\langle N\rangle \approx A \approx M_*
\left(\frac{M_*}{M_P}\right)^2
\approx {\rm TeV}\, ,
\label{mu-term}
\ee
It plays the r\^{o}le of a waterfall field of hybrid inflation,
ending inflation through a phase transition, as discussed in Sec.~\ref{S:inflation}.
And its vev is responsible for generating the right-handed neutrino mass scale.

\subsection{The axion scale}

The most elegant explanation of the strong CP problem is provided by
the Peccei-Quinn (PQ) mechanism~\cite{peccei}, in which the CP
violating angle $\bar\Theta$ is set to zero dynamically as a result of
a global, spontaneously broken $U(1)_{PQ}$ Peccei-Quinn symmetry. The
corresponding Goldstone mode of this symmetry is the axion field and
the static $\bar\Theta$ parameter is substituted by a dynamical one,
$a(x)/f_a$, where $a(x)$ is the axion field and $f_a$ is a
dimensionful constant known as the axion decay constant.

In our model the $U(1)_{PQ}$ Peccei-Quinn symmetry
is spontaneously broken once the scalar fields charged under $U(1)_{PQ}$
(see Eq.  (\ref{PQ})) get a VEV of the order $M_*$. This implies automatically
that the axion decay constant is
\be
f_a \sim M_* \sim 10^{13} {\rm GeV}\,.
\label{fa}
\ee

On the other hand, the axion also has interesting cosmological
implications, especially as a cold
dark matter candidate. Indeed coherent oscillations around the minimum
of its potential may dominate the energy density of the universe if
its potential is very flat. This puts an upper bound for $f_a$ of
order $f_a \leq 10^{12}$ GeV. It seems that our prediction is a little
bit higher that the allowed by experiments. However, as has been
pointed out in~\cite{kim}, $f_a$ can be as big as
$10^{15}$ GeV in models where the reheating temperature is below a
GeV, that is, below the temperature at which the axion field begins to
oscillate. The point is that during inflation the PQ symmetry is
broken and the axion field is displaced at some arbitrary angle, and
it relaxes to zero only after reheating and only below the QCD phase
transition when its potential is tilted. At this point the dangerous
energy stored in the axion field is released, but if the reheating
temperature is of order a GeV then the resulting axion density from
the displaced axion field will be diluted by the entropy release
produced by the inflaton decay. In the
Refs.~\cite{Bastero-Gil:1997vn,leptopaper}
has been showed that
the reheating temperature for the model under consideration is around the GeV
scale and therefore an axion decay constant of the order $10^{13}$ GeV
may be consistent cosmological constraints.

\subsection{The right-handed neutrino mass scale}
Neutrino oscillation phenomenology requires that there must be
two further
heavy {\it right-handed neutrinos} with a Majorana mass arising
from the renormalisable operator ${\lambda_{\nu ,i}}N
(\nu_{R,i}^c)^2$ where ${\lambda_{\nu ,i}}\sim 1$,
and $i=2,3$. We have not
included these operators in the superpotential in Eq.
(\ref{superpotential}) because these heavy right-handed neutrinos
play no r\^{o}le in cosmology, but such operators may readily be
included by suitable choice of PQ charges for the second and third
right-handed neutrinos $\nu_{R,i}$, $i=2,3$. The heaviest right-handed
neutrinos of mass
${\lambda_{\nu ,2}}\langle N\rangle$
and ${\lambda_{\nu ,3}}\langle N\rangle$
will give the dominant
contribution to the solar and atmospheric neutrino masses
of the order of $m_2\approx y_{\nu
,2}^2{v^2}/{\lambda_{\nu ,2}}{\langle N\rangle}$, and
$m_3\approx y_{\nu,3}^2{v^2}/{\lambda_{\nu ,3}}{\langle N\rangle}$,
respectively, where the mild hierarchy $m_2\ll m_3$ can be achieved
by suitable choices of Yukawa couplings above \cite{Antusch:2004gf}.
On the other hand
the {\it lightest right-handed neutrino} which plays an
important r\^{o}le during inflation, and will explain the amplitude
for the curvature perturbation, will play no
part in the see-saw generation of atmospheric and
solar neutrino masses, but will generate the lightest
physical neutrino mass $m_1$.
The lightest right-handed neutrino mass given by
$\lambda_{\nu,1}\langle N\rangle$, where now $\lambda_{\nu,1} \sim
M_*/M_P$, will
contribute to the lightest physical neutrino mass $m_1 =
y_{\nu,1}^2 v^2/(\lambda_{\nu,1} \langle N\rangle)$. A hierarchy
in the neutrino sector
($m_1\ll m_2$) is trivially achieved when
$y_{\nu,1} \ll (M_*/M_P)^{1/2} \, y_{\nu,2} \sim 3\times 10^{-3}
\, y_{\nu,2}$. As we will see in the next section we need such
small Yukawa coupling for $y_{\nu,1}$ in order to stabilize the
preheating effect due to the oscillations of the lightest
right-handed neutrino. Since $m_1$ can be arbitrarily light,
the lightest right-handed neutrino responsible for its mass
can be effectively decoupled from the see-saw
mechanism due to its highly suppressed Yukawa coupling.
The scenario described above is familiar in neutrino phenomenology
and is known as the sequential dominance mechanism \cite{Antusch:2004gf}.

To summarize, a hierarchical neutrino mass scheme, where $m_3\sim 0.05$ eV,
assuming $y_{\nu ,3}\sim 0.1 - 0.5$, led to a right-handed neutrino mass
scale of the same order as the axion scale,
\be
\langle N\rangle \sim M_* \sim 10^{13} {\rm GeV}
\label{nuR}
\ee
Therefore in our model we are led to assign the string scale to the
intermediate scale $M_* \sim 10^{13}$  GeV
by four independent pieces of physics: electroweak symmetry breaking;
the $\mu$ parameter; the axion scale; and the atmospheric neutrino mass scale.

\section{The lightest right-handed sneutrino as a coupled curvaton}
\label{S:inflation}

In a previous paper we suggested that the Higgs fields of the
supersymmetric standard model could play the r\^{o}le of a coupled 
curvaton \cite{Bastero-Gil:2002xr} within this class of models,
and we discussed how Higgs 
perturbations could be converted into the total curvature perturbations
during the first stages of reheating. At first we assumed that
the curvature Higgs contribution does not change after horizon
crossing, and we obtained the desired curvature perturbation
for a Higgs VEV value $h_* \sim 1$ TeV. However we later found that, once
the fields get coupled during the phase transition, the evolution of
the field fluctuations will be affected suppressing the amplitude of
curvature perturbation of the Higgs field relative to its value at horizon
crossing. Subsequently we showed that this suppression 
could be compensated by taking into account 
preheating or parametric resonance effects \cite{preheating}
which can enhance the value of the curvature perturbation
to the desired value \cite{Bastero-Gil:2003tj}.

In this section we propose and study the possibility that the
the r\^ole of the coupled curvaton is instead played by the
lightest right-handed sneutrino, with the inflaton 
identified as the field $\phi$, as before.
The $N$ field and the Higgs fields will here be assumed 
to be zero during inflation.
The associated lightest right-handed
neutrino $\nu_{R,1}$ shall simply be referred to as
$\nu_{R}$, its Majorana Yukawa coupling as $\lambda_\nu =\lambda_{\nu ,1}$,
and its Yukawa coupling to left-handed leptons as $y_\nu = y_{\nu ,1}$,
for ease of notation.

According to the potential in Eq.~\ref{our_potential},
there is a flat direction for both $\tilde{\nu}_{R}$
and $\phi$, while $N$ field is held at zero 
for values of the
inflaton field $\phi$ larger than the critical value
\be
\phi_c = \frac{c_\kappa A + \sqrt{c_\kappa^2 A^2 - 16 m^2}}{4\lambda}
\approx c_\kappa \frac{A}{2\lambda} \, .
\ee
The inflationary epoch is therefore described by a slowly rolling
inflation field $\phi$ and a slowly rolling light right-handed sneutrino field
$\tilde{\nu}_{R}$ (recall that the large mass of the right-handed sneutrino 
is generated by the VEV of the $N$ field, which is zero during 
inflation). As long as
$\phi>\phi_c$, the $N$ field dependent squared mass is positive and
then $N$ (as well as $H$) is trapped at the origin; the potential
energy in Eq.~(\ref{inflation}) is then dominated by the vacuum energy
$V_0$, and the potential (\ref{our_potential}) simplifies
to:
\be
V = V_0 + m_\phi^2 \phi^2 + m^2_\nu \tilde{\nu}_R^2 \,,
\label{inflation}
\ee
with $V_0 \approx F_S^2 \approx (10^8 \rm{GeV})^2$, and $m_\alpha \sim
c_\alpha m$.
The slow roll conditions are given by:
\bea
\epsilon_N &=& \frac{M_P^2 m_\phi^4 \phi_N^2}{V_0^2} < 1 \,,
\label{epsilon}\\
|\eta_N| &=& M_P^2\frac{m_\phi^2}{V_0} <1 \,,
\label{eta}
\eea
where the subscript $N$ means $N$ e-folds before the end of inflation.
Using Eqs.~(\ref{vacuum}) and (\ref{bulkmass}) we then get $|\eta_N| =
c_\phi/c_V $, and $\epsilon_N \ll \eta_N$, and slow-roll only requires
$c_\phi/c_V < 1/3$.

The amplitude of the spectrum of the (comoving) curvature perturbation
${\cal R}$, generated by the inflaton field is given by \cite{lukash}:
\be
P_{\cal R}^{1/2} \simeq
\left(\frac{H_*}{\dot \phi_*}\right)\left(\frac{H_*}{2 \pi}\right)
\simeq \left(\frac{H_*}{2 \pi \eta_N \phi_*}\right) \, ,
\label{cobe}
\ee
where the subscript ``*'' denotes the time of horizon exit, say 60
e-folds before the end of inflation. The value of the inflaton
field  during inflation is around the cutoff of
the theory, $\phi_* \sim \phi_c \sim M_*$, as usual in SUSY
inflation~\cite{riotto}, while the Hubble parameter is of the order of
$H \simeq M_* (M_*/M_P)^3$. Therefore
\be
P_{\cal R}^{1/2} \simeq \left(\frac{M_*}{M_P}\right)^3 \,,
\ee
which for $M_* \simeq 10^{13}$ GeV is quite below the COBE value
$P_{\cal R}^{1/2} = 5 \times 10^{-5}$ \cite{cobe}.

However, the quantum fluctuations of any light field during inflation, i.e.,
the Higgs $h$ and the lightest right-handed
sneutrino $\nu_R$, will contribute to the total
curvature perturbation\footnote{We have dropped the reference to
  the wavenumber in the curvature perturbation but it is implicitly
  assumed that we only refer to
  large scale perturbations, with $k \ll Ha$.}, ${\cal R}$, such that
\cite{gordon,gordon2}
\be
{\cal R} = \sum_\alpha \frac{ \rho_\alpha + P_\alpha}{ \rho+P} {\cal
R}_\alpha \,,
\label{curR}
\ee
$\rho_\alpha$ and $P_\alpha$ being respectively the energy density and
the pressure for each component, with $\rho_\alpha+P_\alpha= \dot
\phi_\alpha^2$, $\rho$ and $P$ the total energy density and pressure,
${\cal R}_\alpha$ the curvature perturbation generated by each
field,
\be
{\cal R}_\alpha \simeq H \frac{Q_\alpha}{\dot \phi_\alpha}
\,,
\ee
and $Q_\alpha$ the gauge invariant quantum fluctuations of the field
\cite{Qi}.
Given the model parameters, we know that the inflaton field has a
background value of the order of the cut-off scale $M_*$, both during
inflation and at the global minimum, but the value of the
sneutrino field is arbitrary during inflation. At the global minimum
it will relax to zero. Given that, we may assume that even
during inflation the sneutrino field is not far from its global minimum
value, and therefore $\phi \gg \nu_R$. From this condition it
follows that $\dot \phi \gg \dot \nu_R$, but ${\cal R}_\phi
\ll {\cal R}_\nu$.
The total curvature perturbation in Eq.  (\ref{curR}) is dominated by the
field with the largest kinetic energy, and during inflation this is just
the inflaton field, which as we have seen gives rise to a too small
contribution.
Nevertheless, the COBE normalization of the
spectrum constraints its value at the onset of the radiation dominated
era, after inflation and the reheating process is complete. In single
field models of inflation, the total curvature perturbation on large
scale remains practically constant after horizon crossing, and it is
enough to estimate the spectrum at that point.
On the other hand, in a multifield inflationary model (or in
general in a multi-component Universe) we have both adiabatic (total) and
entropy or isocurvature perturbations. The latter are given by the
relative contributions between different components, ${\cal R}_\alpha
- {\cal R}_\beta$. Entropy modes can seed the the adiabatic one, i.e.,
the total curvature perturbation, when their contribution to the total
energy density becomes comparable
\cite{silvia,mukhanov,polarksi,deruelle,juandavid,riotto2}.
This is what we expect at the end
of inflation, when all the fields move fast toward the global
minimum. At this point, the energy densities of the fields become
comparable, and the total curvature perturbation,
Eq.  (\ref{curR}) may become of the order of ${\cal R}_\nu$.

Here we would like to consider instead the lightest
right-handed sneutrino as the main
source of the isocurvature perturbation during inflation. 
The Yukawa coupling
of the lightest right-handed sneutrino to the $N$ field is a
factor $(M_P/M_*)$ larger than the coupling $\kappa$ between $\phi$
and $N$. Hence, the right handed sneutrino acquires a mass in the global
minimum which is larger than the other particles by the same factor.
Nevertheless, due
to the coupling with the other fields, it may
oscillate together with $N$ and $\phi$ but with a smaller amplitude
$\nu_R \sim (\kappa/\lambda_\nu) M_*$. In any case, during the oscillations
the energy density of the 3 fields become comparable, with $M_{RR}\,
\nu_R \sim (\kappa N) N$.

Previous to that we have to consider the effect of the phase
transition on the perturbations of the fields. The tachyonic
instability in the $N$ direction makes the field grow until it
reaches the straight-line trajectory in field space, $N= \sqrt{2} (
\phi - \phi_c)$. On the other hand the sneutrino still follows
undisturbed its inflationary trajectory for a while. With respect to
$Q_N$ perturbations, they also feel the tachyonic instability, they
grow as $Q_N \propto \dot N$ and we end up with $Q_N = -\sqrt{2} Q_\phi$
along the straight-line trajectory. At this point no change has
been induced in the total curvature perturbation. The fact that $Q_N$
is non negligible after the phase transition does not mean that we
are generating additional entropy modes. Along the straight line
trajectory we still have only one degree of freedom, the adiabatic
mode \cite{gordon2}.

Soon after that, the lightest right-handed sneutrino starts feeling
the presence of the
other fields. The effect on its perturbation is the same
than we found for the Higgs in Ref. \cite{Bastero-Gil:2003tj}: when the field
gets coupled the ${\cal R}_\nu$ perturbation is dragged toward
${\cal R}$, and by the time they first reach the global minimum all
the contributions ${\cal R}_\alpha$ are comparable but some orders of
magnitude smaller than the initial ${\cal R}_\nu$.

From this point of view, during the phase transition the initial
entropy perturbation is only partially converted into the
adiabatic one. In other words, entropy perturbations are suppressed
due to the tachyonic instability during the phase transition.  And in
particular for our model values, we are still
some orders of magnitude below the COBE normalization. Again, once the
oscillations begin, the presence of a third field, in this case the
sneutrino, will curve the initial straight-line trajectory in the
$N$-$\phi$ plane, giving rise to preheating of the large scale
perturbations. In this scenario this will happen before the fields
reach the global minimum. We have the inflaton field decreasing from
its value at the critical point, while $N$ and $\tilde \nu_R$ are
growing and moving faster than $\phi$. First $N$ gets destabilized,
and then $\tilde \nu_R$. But for the values of the
fields such that  $\lambda_\nu \tilde \nu_R^2/M_* >  \kappa (\phi_c
-\phi)$, there is an approximate minimum with $N \simeq \tilde \nu_R$
and $\lambda_\nu \tilde \nu_R^2/M_* \simeq \phi \simeq \phi_c$, around which
$N$ and the RH sneutrino oscillates. It is already during this period
that the large scale perturbations are preheated. First,
$N$ and $\tilde \nu_R$ do not oscillate in phase, and in addition
there is a tachyonic instability in the $N$ and $\tilde \nu_R$
squared masses  during the oscillations, which enhances the resonance
\cite{juantachyonic}. This effect only lasts a short period of time,
until the inflaton field is close enough to the global minimum to drag
towards it the other fields.

At the same time we may also preheat fluctuations on
smaller scales\footnote{Without the sneutrino and/or the Higgs, there
is indeed a strong parametric resonance (tachyonic preheating
\cite{juantachyonic}) for the fluctuations with a
wavenumber up to $O(\kappa N_0)$ \cite{prehhybrid} during the first 3
oscillations of the fields, before it enters in a narrow resonance
regime.}, which will back-react on the system shutting down the
resonance. But we do not expect this to happen until the inflaton is
also oscillating, due to the small amplitude of the previous
$N-\tilde \nu_R$ oscillations. Another effect to be considered is the
decay of the heavy sneutrino, which happens earlier than the decay of the
singlets $\phi$ and $N$. Once the lightest right-handed sneutrino (the
source of the entropy perturbation) disappears  the resonance on the
large scales will stop.

\begin{figure}[t]
\begin{tabular}{cc}
\epsfysize=10cm \epsfxsize=8cm \hfil \epsfbox{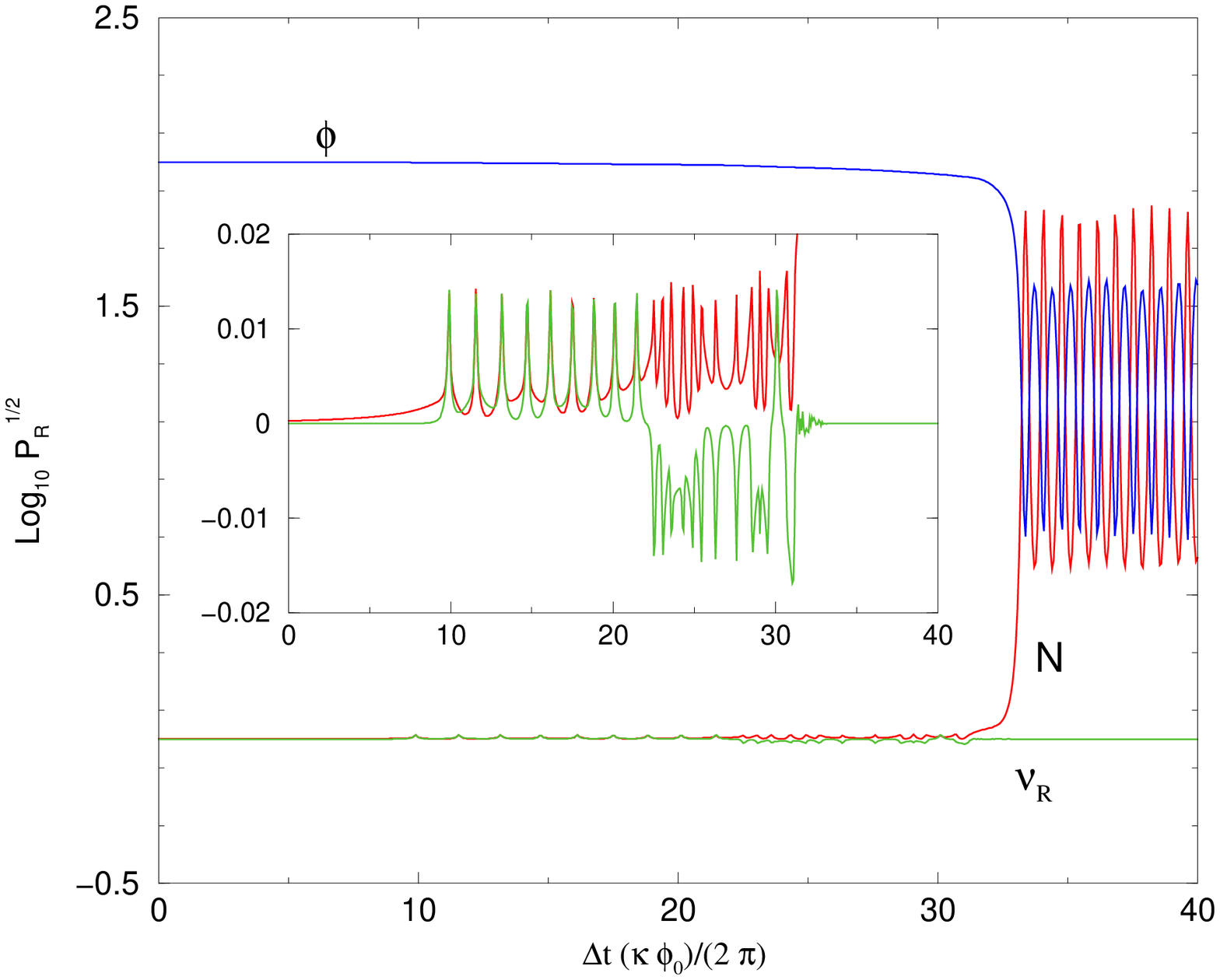} \hfil
\epsfysize=10cm \epsfxsize=8cm \hfil \epsfbox{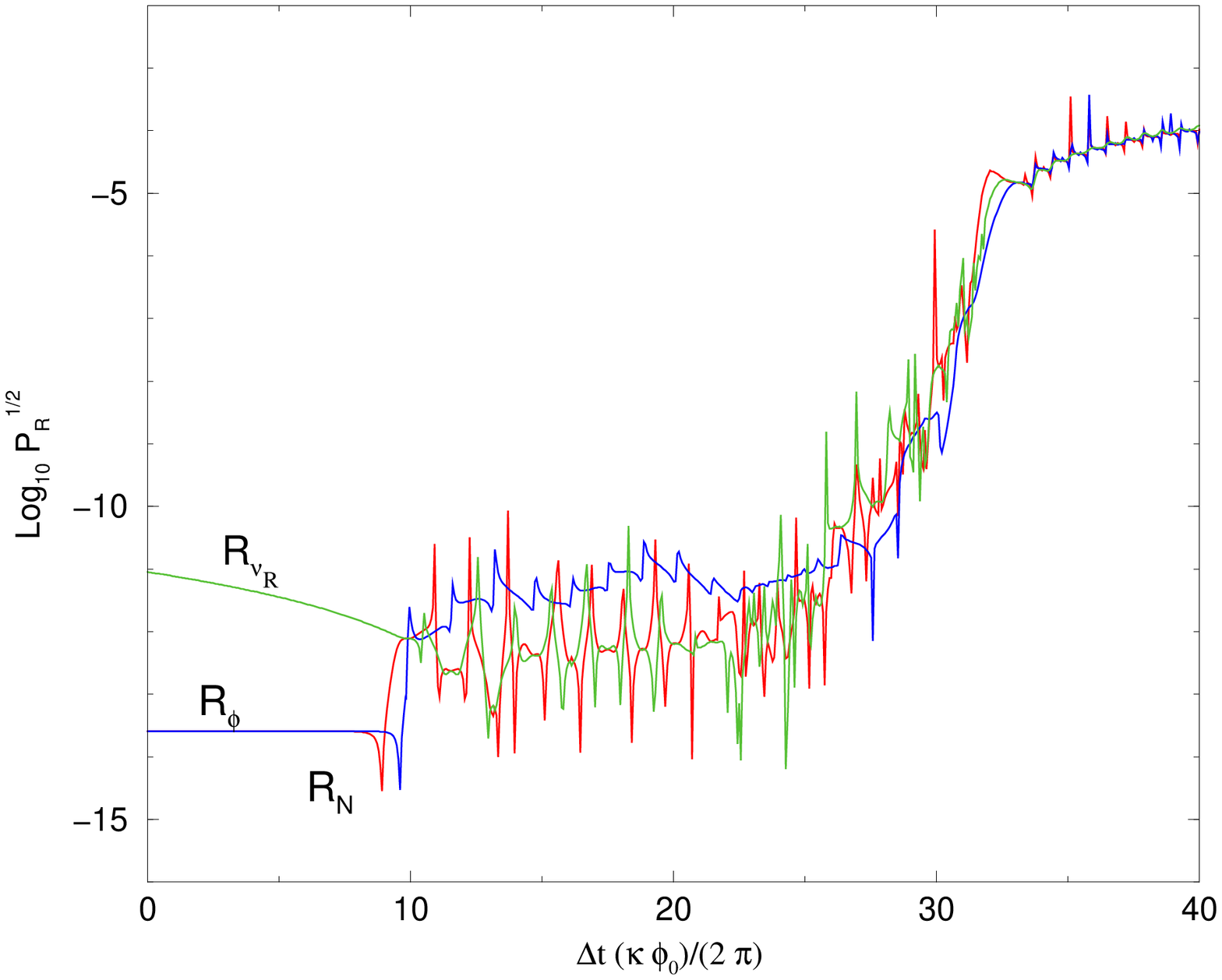} \hfil
\end{tabular}
\caption{\footnotesize Left plot: Evolution of the background fields,
$\phi$, $N$, and $\tilde \nu_R$, after they have passed the critical
point; inset: oscillations of $N$ and $\tilde \nu_R$, previous to that
of $\phi$. Right plot:
Amplitude of the spectrum of the curvature
$P^{1/2}_{{\cal R}_\alpha}$ for the different components of the model:
inflaton $\phi$, $N$ field and lightest right-handed sneutrino
sneutrino $\tilde{\nu}_R$.
\label{fig1}}
\end{figure}

As an example, in Fig. (\ref{fig1}) we have plotted the evolution of
the background fields (LHS plot), and the different
field contributions to the curvature perturbation ${\cal R}_\alpha$
(RHS plot),
during the first oscillations of the fields, after they have passed
the critical point. The value of the lightest right-handed
sneutrino field during
inflation is $\nu_R = 1$ TeV, which gives ${\cal R}_{\nu*} \simeq
10^{-5}$ at horizon-crossing. Nevertheless, as it can be seen in the
plot, after the phase transition all contributions are roughly equal.
For the numerical integration, we have chosen
$\lambda_\nu = 2 \times 10^{5} \kappa$, and we have also
introduced a decay rate for the lightest right-handed
sneutrino, $\Gamma_\nu \simeq
y_\nu^2 M_{RR}$, with $y_\nu \simeq 7\times 10^{-3}$. This gives rise to the
right order of magnitude for the amplitude of the curvature
perturbation (slightly larger than COBE).
The main enhancement is produced
during the oscillations of the fields $N$ and $\tilde \nu_R$, and the
resonance ends when $\tilde \nu_R$ goes to zero and we are left only
with $\phi$ and $N$ oscillating.

The decay of the lightest right-handed
sneutrino gives rise to radiation, with $\rho_R
\simeq 0.2 \rho_\phi$, where $\rho_\phi$ refers to the energy density
in the oscillating singlets. 
\footnote{The decay of the lightest right-handed sneutrino
is also expected to generate leptogenesis, 
and cold dark matter can arise from preheating of neutralinos 
as discussed in \cite{Bastero-Gil:2000jd}.}
Having converted a fraction of the initial vacuum
energy into radiation, the background fields still oscillate (in
phase) but with smaller amplitudes compared to the case without the
sneutrino, which may not be large enough to allow the preheating of
the small scale fluctuations.
Nevertheless one has to bear in mind that larger decay
rates for the sneutrino means a shorter resonance for the curvature
perturbation, or no resonance at all. On the other hand, for a smaller
decay rate the resonance will then be shut down by the backreaction of
the small scale modes. 

\section{Conclusion}

We have proposed a model of cosmology and particle physics
in which all relevant scales are derived from an
intermediate string scale $M_*\sim 10^{13}$ GeV, identified with
both the Peccei-Quinn symmetry breaking axion scale $f_a$
and the heaviest right-handed neutrino mass scale $M_{RR}$.
A supersymmetry breaking scale is derived from the
constraint of having a small cosmological constant
leading to $F_S^{1/2}\sim M_*^2/M_P\sim 10^8$ GeV.
The $\mu$ parameter of the MSSM and the electroweak breaking scale are then
given by $F_S/M_*\sim 10^3$ GeV.
The model involves hybrid inflation, with the inflaton
mass given by $F_S/M_P \sim $ MeV, and their Yukawa couplings given by
$(M_*/M_P)^2\sim 10^{-10}$.
In our model we were led to assign the string scale to the
intermediate scale $M_* \sim 10^{13} {\rm GeV}$
by four independent pieces of physics: electroweak symmetry breaking;
the $\mu$ parameter; the axion scale; and the neutrino mass scale.
In recent years models with an intermediate string scale
($10^{11}<M_*<10^{14}$ GeV) have been seen of great interest because
contain many phenomenological issues (see for example
\cite{benakli,abel}). The novelty of our construction is we give an
explicit potential (\ref{our_potential}) where the link between
particle physics and cosmology is explicit. Also the intermediate
scale appears as a consequence of requiring zero cosmological constant 
at the global minimum of the potential together with requiring
one of the mass parameters involved in the
potential to be the electroweak scale.
In addition, if all the dimensionless couplings are
equal in the higher dimensional model, we found that the effective
couplings are just powers of the ratio $M_*/M_P$.

The large scale structure in the
Universe is generated by the lightest right-handed sneutrino
playing the r\^{o}le of a coupled curvaton.
We showed that it is possible to obtain the correct
curvature perturbation by including the effects of preheating and
adjusting neutrino Yukawa couplings to be rather small,
leading to a (lightest) right-handed neutrino mass of about
$10^8$ GeV. This is possible because of sequential dominance
\cite{Antusch:2004gf} since in this case the lightest right-handed neutrino
plays no essential r\^{o}le in the see-saw mechanism.
The sneutrino as a coupled curvaton only strengthens the
synthesis between cosmology and particle physics at the
intermediate scale.

\section*{Acknowledgements}
VDC was supported at the University of Oxford by PPARC rolling grant PPA/G/0/2002/00479.

\section*{Appendix A: Yukawa Couplings}
In higher dimensions the superpotential is well defined just in one of the
fixed points (Brane I or Brane II)~\footnote{Notice that this is
always the case if the coupling involves bulk and brane fields. However by
construction also couplings which involve just bulk fields have to be
defined in one of the four dimensional boundary due to an enhancement of
the number of supersymmetries contained in the bulk.}.
In general the Lagrangian in higher dimensions is given by
\be
{\cal L}_{4+d} = \int d^2\theta\,\, \hat W_{4+d} \left(\delta^d(0) +
\delta^d(y_i-\pi R)\right) \, ,
\label{eq.1}
\ee
where the superpotential $\hat W_{4+d}$ is a function of bulk fields ($\hat
\Psi_i$) and brane fields ($\Phi_j$), given by:
\be
W_{4+d} \ni \frac{\hat \lambda_{i,j}}{M_*^{\frac{\alpha d}{2}}}
\frac{\hat\Psi_i^\alpha \Phi_j^\beta}{M_*^{\alpha+\beta-3}} \, .
\label{W4d}
\ee
Here we have introduced a mass scale, $M_*$, in such a way the couplings
$\hat \lambda_{i,j}$ remain dimensionless. This scale is actually the
Planck scale in higher dimensions, or the string scale in string theories,
which is related with the four dimensional
Planck scale, $M_P$, through the well known formula
\be
M_P^2 = M_*^{2+d} R^d \, .
\label{planck}
\ee
Using the fact that the bulk fields contain a volume suppression factor
with respect their zero mode (the effective four dimensional field), i.e,
$\hat \Psi_i = \Psi_i /R^{d/2}$, we can express the four dimensional
effective superpotential, after integrating out the extra $d$ dimensions, as
\bea
W_4 &=& \int d^d y \,\,\hat W_{4+d} \left(\delta^d(0) + \delta^d(y_i-\pi
R)\right) \nonumber \\
&=& \hat \lambda_{i,j} \left(\frac{M_*}{M_P}\right)^\alpha
\frac{\Psi_i^\alpha \Phi_j^\beta}{M_*^{\alpha+\beta-3}} \, .
\label{W4}
\eea
Redefining the effective four dimensional coupling as
\be
\lambda_{i,j} = \left(\frac{M_*}{M_P}\right)^\alpha \hat \lambda_{i,j} \, ,
\label{couplingd}
\ee
we notice that in general we get an important suppression factor if
$M_* \ll M_P$~\footnote{This suppression factor only would depend on the
number $\alpha$ of fields living in the bulk and it will be completely
independent of the number $d$ of extra dimensions.}.
In other words, if there is a fundamental intermediate scale, $M_*$,
defined in the higher dimensional theory, we always get small couplings,
$\lambda_{i,j} \ll 1$, even though the higher dimensional couplings,
$\hat\lambda_{i,j}$, can be of order one or even non-perturbative as it
would be in our particular case.

Finally note that following the same procedure in the
Kahler potential will lead to no additional volume effects.
For example consider a canonically normalized $4+d$-dimensional Kahler potential
term of the form $\hat\Psi_i\hat\Psi_i^\dagger$ then the corresponding
$4$-dimensional Kahler potential will contain the term $\Psi_i\Psi_i^\dagger$
which maintains its canonical form due to a cancellation of the volume factors.
There may be some small additional corrections due to canonical normalization
effects \cite{King:2004tx}, but these will not affect the analysis here.

\section*{Appendix B: Soft Masses}
We shall suppose that SUSY is broken by the $F_S$-term of a four dimensional
gauge-singlet field $S$ localized at the Brane II ($y_i = \pi R$) and
mediated across the extra dimensional space to the Brane I by bulk
fields propagating in a loop correction like gaugino mediation~\cite{gaugino}.
The SUSY breaking Lagrangian contains six terms that lead to the bulk gaugino
mass ($M_{A}$), Brane II gaugino mass ($M_{B}$),
vacuum energy ($V_0$), soft masses for bulk fields ($M_{\Psi_i}$), soft masses
for
Brane II fields ($M_{\Phi_i}$) and trilinear soft terms
($A_{\lambda_{i,j}}$). In general we have
\bea
{\cal L}_{4+d}^{soft} &=& \delta^d(y_i-\pi R) \left ( c_{\lambda_{ij}} \int d^2 \theta
\,\,\frac{\hat W_{4+d} S}{ M_*}  + c_{\Psi_i}\int d^4\theta \,\, \frac{\hat\Psi_i^\dagger
\hat\Psi_i S^\dagger S}{M_*^{2+d}} +  c_{\Phi_i}\int d^4 \theta \,\,
\frac{\Phi_i^\dagger \Phi_i S^\dagger S}{M_*^2} \right. \nonumber \\
&+& \left. c_V \int d^4 \theta \,\,S^\dagger S +
c_{A}\int d^2\theta\,\, \frac{\hat W_\alpha^{(A)} \hat W^{\alpha(A)}
S}{M_*^{1+d}} + c_{B}\int d^2\theta\,\, \frac{W_\alpha^{(B)}
W^{\alpha(B)} S}{M_*} \right) \, ,
\label{susy}
\eea
where $W_\alpha^{(A)}$ ($W_\alpha^{(B)}$) is the field strength of the gauge
group $G_A$ ($G_B$) and the constants $c_i = \{c_{\lambda_{ij}}, c_{\Psi_i}, c_{\Phi_i},
c_V, c_{A}, c_{B } \}$ are of the order
one. These constants are completely model-dependent, for example, the
non-renomalizable terms (\ref{susy}) might come from integrating out some modes
of
mass $M_*$ propagating in a loop process such that $c_i ={\cal
N}/(4\pi)^2$, being $\cal N$ the massive mode's degree of freedom. It
is straightforward to show that in the effective four dimensional theory we
get different mass scales asociated with bulk fields and branes fields as
along as $M_* \ll M_P$. These are given by
\bea
V_0 &=& c_V F_S^2 \label{vacuum} \\
M_A &=& c_{A} \frac{F_S}{M_P} \hspace{1cm}  M_{\Psi_i}^2 = c_{\Psi_i}
\left(\frac{F_S}{M_P}\right)^2 \label{bulkmass}\\
M_B &=& c_{B} \frac{F_S}{M_*} \hspace{1cm}  M_{\Phi_i}^2 = c_{\Phi_i}
\left(\frac{F_S}{M_*}\right)^2 \hspace{1cm} A_{\lambda_{i,j}} = c_{\lambda_{ij}}
\frac{F_S}{M_*} \, .
\label{branemass}
\eea
The vacuum energy ($V_0$) would
dominate the total energy density during inflation providing the typical
expansion rate (the Hubble constant) as~\footnote{As we have said at the
beginning of the section one important assumption of our model is that the
location of the two branes are already stabilized before inflation takes
place wich means that only the four dimensional space are inflated away. In the
case that all de $4+d$-dimensions feel inflation at the same time, the
Hubble constant would be given by $H^2 = V_0/(3 M_*^2)$ instead~\cite{arkani}.}
\be
H^2 = \frac{V_0}{3 M_P^2} = \frac{c_V}{3} \left(\frac{F_S}{M_P}\right)^2
=
\frac{c_V}{3c_{\Psi_i}}  M_{\Psi_i}^2 \, .
\label{hubble}
\ee
The bulk mass scale ($M_{\Psi_i}$) would give us, for example, the inflaton
mass. Therefore, in order to satisfy the slow roll condition during
inflation ($H<M_{\Psi_i}$) it turns out from (\ref{hubble}) that we just
need some tuning on the constants $c_i$, i.e. $c_V <
3 c_{\Psi_i}$. Finally, the Brane II mass scale would define the typical MSSM
soft
term, $M_{SUSY} \sim M_{\Phi_i} \sim {\cal O}(\mbox{TeV})$. Notice that in
general we would have $H\ll M_{SUSY}$ which is completely different what
happen in the normal four dimensional supersymmetric hybrid inflationary
models where $H\sim M_{SUSY}$~\cite{riotto}.


\end{document}